\documentclass[trackchanges]{aastex701}
\usepackage{booktabs}
\usepackage{multirow}
\usepackage{ulem}

\begin{document}

\title{X-ray emission of F-type stars and its analogy with G-type stars}

\author[]{Takuma Shimura}
\affiliation{Graduate School of Science, Tokai National Higher Education and Research System, Nagoya University, Nagoya, Aichi 464-8602, Japan}
\email{shimura_t@u.phys.nagoya-u.ac.jp}  

\author[orcid=0000-0002-9901-233X]{Ikuyuki Mitsuishi} 
\affiliation{Graduate School of Science, Tokai National Higher Education and Research System, Nagoya University, Nagoya, Aichi 464-8602, Japan}
\email[show]{mitsuisi@u.phys.nagoya-u.ac.jp}  

\author[orcid=0009-0005-2657-2554]{Koki Sakuta} 
\affiliation{Astronomical Institute, Tohoku University, 6-3 Aramakiazaaoba, Aoba-ku, Sendai, Miyagi 980-8578, Japan}
\email{koki.sakuta.a1@tohoku.ac.jp}  

\author[orcid=0000-0002-1932-3358]{Masanobu Kunitomo}
\affiliation{Department of Physics, Kurume University, 67 Asahimachi, Kurume, Fukuoka, 830-0011, Japan}
\email{kunitomo.masanobu@gmail.com}

\author[orcid=0000-0003-3882-3945]{Shinsuke Takasao}
\affiliation{Humanities and Sciences/Museum Careers, Musashino Art University, Kodaira, Tokyo 187-8505, Japan}
\email{stakasao@musabi.ac.jp}

\author[orcid=0000-0002-0141-5131]{Yuki A. Tanaka}
\affiliation{National Institute of Technology, Fukushima College, Iwaki, Fukushima 970-8034, Japan}
\email{ytanaka@fukushima-nct.ac.jp}



\begin{abstract}

We conducted a systematic spectral study for X-ray bright single F-type main-sequence (MS) stars without significant X-ray outbursts to investigate X-ray spectral properties 
such as temperature, emission measure (EM), and luminosity ($L_{\rm X}$).
To ensure sufficient X-ray photon statistics, 33 single stars were selected by cross-matching the XMM-Newton source catalog and the Tycho-2 spectral type catalog, although the resulting sample is not statistically complete.
A positive correlation was found in the observed EM--EM-weighted temperature relationship as seen in late-type stars.
Our single F-type MS stars have coronal plasma with an EM-weighted temperature of $\lesssim$1 keV and 
an EM of $\lesssim$10$^{53}$ cm$^{-3}$ corresponding to $L_{\rm X}$ of $\lesssim$10$^{30}$~erg~s$^{-1}$.
These observational features for the single F-type MS stars are consistent with those of the single G dwarf stars, 
suggesting that their overall X-ray coronal properties are broadly similar.
The column EM, defined as the total EM divided by the stellar surface area, also shows broadly similar distributions for single F- and G-type stars, although some G-type stars exhibit higher column EM values, possibly reflecting differences in coronal magnetic activity.
Additionally, the obtained relationship between the X-ray activity and the Rossby number reinforces this suggestion in the literature. 
Moreover, the upper bounds in EM and $L_{\rm X}$ were found to be characteristic of single stars in our sample and not valid for binary stars.
Our results suggest that the planetary evolution in terms of the X-ray properties around F-type MS stars can be understood by extending the frameworks developed for G-type stars.
\end{abstract}

\keywords{stars: coronae --- stars: activity --- stars: solar-type --- X-rays: stars --- Sun: X-rays, gamma rays}


\section{Introduction} \label{sec:intro}

It has been recognized that stellar high-energy radiation such as X-rays can affect the physical and chemical structures of planetary atmospheres and their evolution \citep[e.g.,][]{Airapetian+16, Linsky19, Owen19}. High-energy radiation such as X-rays originates from stellar coronae. Considering the long lifetime of the main-sequence (MS) phase, understanding the X-ray properties of MS stars, or stellar coronal properties, is therefore important for revealing planetary evolution. The stellar coronal properties that should be revealed are not only the X-ray luminosity but also the coronal temperature, which determines the hardness of the X-ray spectrum. To date, such properties have been mainly investigated for late-type stars such as M-, K-, and G-type stars \citep[e.g.][]{gudel1997,Johnstone2015,2020ApJ...901...70T}. However, recently many planets have also been detected around other spectral types of stars such as more massive, rare F-type MS stars \citep[e.g.][]{1997ApJ...474L.115B,2014A&A...562L...3G}, increasing the importance of coronal studies of other types of MS stars.

X-ray luminosities have been investigated in many previous systematic studies for a variety of spectral types \citep[e.g.][]{1996ApJS..102...75M,1997A&A...318..215S,1998A&AS..132..155H,1999A&A...341..751M,2002A&A...383..210M,2003SSRv..108..577F,2007A&A...475..677S,2012MNRAS.422.2024J,2018A&A...620A..89N,2019A&A...628A..41P}; however, without detailed spectral analysis.
They reported several noteworthy findings, including a positive correlation between the total emitted X-ray surface flux and spectral hardness, as well as evidence supporting the traditional distinction between the saturated and correlated regimes in the X-ray activity--rotation relation, suggesting the presence of common observational features across spectral types.
%
%
Notably, the number of systematic studies (with detailed spectral analysis) extracting the metal abundance, the hydrogen absorption column density, the temperature ($kT$), the emission measure (EM), and the luminosity \citep[e.g.][]{gudel1997,2011ApJS..194....7N,Johnstone2015,2019A&A...628A..41P,2020ApJ...901...70T} is relatively small.
The spectral analysis allows us to compare theoretical predictions with observations. 
\citet{2020ApJ...901...70T} derived a theoretical EM--EM-weighted temperature ($\bar{kT}$) scaling law for stellar coronae. 
The model assumes that the stellar X-ray radiation is the sum of the contribution from multiple active regions. 
They adopted the size distribution function of magnetic elements based on solar observations \citep{2009ApJ...698...75P}. 
It is also assumed that the thermal property of each active region can be described by the Rosner-Tucker-Vaiana scaling law \citep{Rosner1978,Shibata2002}.
They compared the scaling laws to the observational data of the single G dwarf stars including the Sun and 
found that with the solar parameters, the scaling law seems to be consistent with the data of slowly rotating stars.
The scaling law may explain the observations of (possibly rapidly rotating) X-ray-bright stars 
if these stars show a size distribution function of active regions with the same power-law index as the solar one but a 10--100 times larger coefficient.
In this study, we extend our previous study on G dwarf stars \citep{2020ApJ...901...70T} to include F-type MS stars. 
F-type MS stars with surface convection layers may have coronal properties similar to those of G dwarf stars because their stellar masses and radii are relatively similar. However, their convection zones are believed to be thinner, and the convective turnover time ($\tau_{\rm c}$) is expected to change dramatically across the F0--F9 spectral range \citep{2011ApJ...741...54C}.
%
In previous studies, for example, \citet{2007A&A...475..677S} found that 
the overall {\it ROSAT} detection rate of bright A-type stars lies between 10--15\% in the spectral range A0--A9, 
while a steep increase in detection rates was seen among F-type stars, suggesting that the onset of the dynamo action is clearly visible.
\citet{2019A&A...628A..41P} noted that F-type stars seem not to participate in the downward trend of the X-ray emission in the rotation period (or Rossby number) range observed for later spectral types and show no clear correlation between X-ray luminosity and UV excess luminosity. They concluded that it is not obvious to attribute this finding to the transition to a fully radiative interior with the ensuing breakdown of the solar-type dynamo. They also proposed that close binary systems with later-type companion stars may partly explain the unusual activity behavior.
%
%
However, previous observations are insufficient to quantitatively assess the similarities and differences in coronal properties between F-type and other stars because of the lack of detailed spectroscopic data. Such data are essential for comparing the thermal properties of stellar coronae.
%
%
In addition, distinguishing single and binary stars is crucial for X-ray data analysis and we limit our analysis to single stars.
%
%

%
The rest of the paper is organized as follows: Section 2 describes the sample selection, observations, and data reduction. Section 3 presents our analysis methods and results. 
Section 4 discusses the X-ray spectral properties of single F-type MS stars, focusing on the EM--$\bar{kT}$ relation and the relationship between X-ray activity and the Rossby number. 
Finally, Section 5 summarizes our conclusions.
 
%
%
%
%

%

\section{Observations and data reduction} \label{sec:obs_datareduction}
To systematically investigate the X-ray spectral properties of stellar plasmas 
of F-type MS stars, we need a large sample. 
As is the case with our previous study \citep{2020ApJ...901...70T}, we cross-matched two large astronomical catalogs, 
the Tycho-2 spectral type catalog \citep{2003AJ....125..359W} and the 4XMM-DR10 catalog \citep[e.g.][]{2020A&A...641A.136W} with a matching radius of 10 arcseconds. 
Both of them have a large number of registered sources. 
The Tycho-2 spectral type catalog contains approximately 350,000 stars with detailed information (e.g., spectral type and luminosity class) for each star. 
The 4XMM-DR10 catalog is the catalog based on the observation data of the {\it XMM-Newton} satellite, 
which contains more than 800,000 X-ray sources detected in the field of view of more than 10,000 observations. 
In addition to the coordinates, the catalog contains various information (e.g., X-ray fluxes, extent, variability, and the number of X-ray photons observed per detector) useful for this research. 
For most of the cross-matched sources, the separation angle between the two catalogs is within 3 arcseconds. 

For the cross-matched star candidates, we selected those with an extent parameter of 0 arcseconds. 
Then, to obtain a suitable sample for spectroscopic analysis, we selected the sources 
with a source photon count of more than 400 per one EPIC detector after the background correction tabulated in the 4XMM-DR10 catalog. 
We confirmed that the spectral type of the candidates is classified as an F-type MS star in the SIMBAD astronomical database 
and the nominal values of the effective temperature in the {\it Gaia} DR2 database \citep[e.g.][]{2016A&A...595A...1G} are included between 6000 and 7500 K. 
Next, to exclude binary systems, we referred to the SIMBAD astronomical database and 
several binary catalogs \citep{2002A&A...384..180F,2004A&A...424..727P,2019A&A...623A..72K,2021MNRAS.506.2269E} 
to ascertain that our sample is not classified into a binary system in their catalogs. 
Finally, to improve the quality of our spectral analysis, we excluded stars having any additional corresponding sources tabulated in the SIMBAD database and the 4XMM-DR10 catalog within a radius of 30 arcseconds.
%
Explicit proper-motion corrections were not applied in the present analysis. However, we carefully checked the positional consistency within the adopted 10 arcsec matching radius and excluded ambiguous cases with possible nearby contamination.
Additionally, if a spectral analysis area of a candidate overlaps that of a diffuse source with a non-zero extent parameter, the candidate was removed from our final target list.
As a result, considering the spectral analysis results described in the next section, a total of 33 sources were selected after removing 14 targets.
%

Information regarding the stars used in this study is summarized in Table \ref{tab:obssummary}. 
We have used HEAsoft v6.21 and XSPEC version 12.9 and, if not otherwise specified, 
the error ranges show the 90\% levels of confidence from the central values.
Thirty four data sets were obtained from the science archive of the {\it XMM-Newton} satellite. 
To conduct high-quality spectroscopic analysis, data reduction was performed by using SAS 18.0.0, 
which is the standard software designed to reduce and analyze the {\it XMM-Newton} data,  
as is the case with our previous study \citep{2020ApJ...901...70T}.
We used the data obtained by the PN detector, which has the largest effective area among the EPIC detectors and 
therefore is expected to collect the largest number of photons.
The original data were used to obtain photon statistics, except in cases where significant discrepancies were observed in the background-subtracted energy spectra before and after the screening.
%
Next, to investigate the spectroscopic characteristics of non-flaring (or quasi-steady) plasmas, 
we investigated the source variability for each target 
and confirmed that the source variability flags tabulated in the X-ray catalog are set to be false for all the targets. 
%
%
Additionally, we created light curves in 300--600 and 600--1500 eV and 
extracted hardness ratios of both energy bands which are believed to be sensitive 
to the spectral shape for hot stellar plasmas with a time bin of at least 500 seconds. 
We confirmed that there are no significant flare-like features during the observation time used in our spectral analysis for our final target list.
For HD 36364, two observations were found and their source variability flags were set to be false in the catalog.
We confirmed that the resultant parameters are consistent with each other within the statistical error and 
we used the results for the 0763720401 observation with higher photon counts in the discussion. 
%

\begin{table*}[h!]
 \caption{Basic information and logs of observations for F-type MS stars.}
  \begin{center}
\begin{tabular}{llllllllll}
\hline \noalign{\vskip2pt} 
Target Name & Target ID$^a$ & R.A.$_{{\rm J2000}}$     &   Dec.$_{{\rm J2000}}$   & SpT$^b$   & T$_{\rm{eff}}$$^c$ & dist.$^c$ &  Obs. ID$^d$     &  Exp.$^e$  &  Counts   \\
&           & [deg.]     & [deg.]                    &  & [K]   &  [pc]  &  &  [ks]    & \\ \hline
HD 224997 & 4XMMJ000224.6-060334 & 0.603 & -6.060 & F5V & 6287 & 130 & 0652010401 & 27 & 453 \\ \hline
HD 5677 & 4XMMJ005802.6-361201 & 14.511 & -36.200 & F0V & 6633 & 234 & 0205680101 & 83 & 433 \\ \hline
HD 7779 & 4XMMJ011456.5-734622 & 18.735 & -73.773 & F3/5V & 6606 & 233 & 0784690301 & 34 & 451 \\ \hline
HD 13456 & 4XMMJ021122.2-100309  &  32.843 & -10.053 & F3V & 6700 & 50 & 0204340201 & 9 & 1310 \\ \hline
HD 14224 & 4XMMJ021751.0-042122 & 34.462 & -4.356 & F3V & 6416 & 136 & 0785100101 & 17 & 440 \\ \hline
HD 14245 & 4XMMJ021804.2-035014 & 34.518 & -3.837 & F7/8V & 6076 & 116 & 0760540101 & 103 & 3068 \\ \hline
HD 15557 & 4XMMJ023239.8+614403 & 38.166 & 61.734 & F0V & 7019 & 115 & 0692080201 & 39 & 677 \\ \hline
HD 23352 & 4XMMJ034524.1+245308 & 56.351 & 24.886 & F5V & 6145 & 135 & 0761920501 & 72 & 896 \\ \hline
HD 32848 & 4XMMJ050113.8-653309 & 75.308 & -65.553 & F5V & 6339 & 99 & 0149630301 & 21 & 1658 \\ \hline
HD 32645 & 4XMMJ050421.8-081402 & 76.091 & -8.234 & F3/5V & 6564 & 116 & 0554700301 & 22 & 1786 \\ \hline
HD 36364 & 4XMMJ053100.6-043557 & 82.753 & -4.599  & F2V & 6419 & 328 & 0763720401    &   93 & 1039 \\ \hline
HD 37552 & 4XMMJ053351.2-675621 & 83.464 & -67.939 & F5V & 6517 & 120 & 0671010401 & 22 & 453 \\ \hline
HD 37916 & 4XMMJ054135.9-092935 & 85.400 & -9.493 & F2V & 6895 & 101 & 0740570101 & 43 & 1081 \\ \hline
HD 39904 & 4XMMJ054928.7-694715 & 87.370 & -69.788 & F2V & 6618 & 132 & 0690751301 & 22 & 709 \\ \hline
BD+19 2050 & 4XMMJ083728.1+190944 & 129.367 & 19.162 & F6V & 6637 & 175 & 0761921101 & 51 & 800 \\ \hline
BD+20 2176 & 4XMMJ084110.0+193031 & 130.292 & 19.509 & F7V & 6234 & 186 & 0742570101 & 55 & 819 \\ \hline
BD+20 2180 & 4XMMJ084126.9+193232 & 130.362 & 19.542 & F4V & 6466 & 180 & 0742570101 & 55 & 798 \\ \hline
HD 73937 & 4XMMJ084136.1+190833 & 130.401 & 19.143 & F2V & 7018 & 195 & 0761921001 & 25 & 787 \\ \hline
HD 104456 & 4XMMJ120141.4-184727 & 180.423 & -18.791 & F3/5V & 6371 & 126 & 085220201 & 36 & 458 \\ \hline 
HD 114746 & 4XMMJ131242.4-192643 &  198.177 & -19.445 & F5V & 6007 & 236 & 0502070101 & 101 & 927 \\ \hline
HD 117822 & 4XMMJ133318.0-315302  & 203.325 & -31.884 & F3V & 6629 & 140 & 0105261301 & 38 & 400 \\ \hline
HD 119718 & 4XMMJ134635.3-620409 & 206.647 & -62.069 & F5V & 6386 & 115 & 0742110101 & 26 & 750 \\ \hline
HD 122390 & 4XMMJ140152.7-112822 & 210.470 & -11.473 & F5V & 6223 & 109 & 0610980201 & 30 & 1391 \\ \hline
HD 150345 & 4XMMJ164126.5-242409 & 250.361 & -24.403 & F5V & 6379 & 81 & 0112480201 & 12 & 1994 \\ \hline
HD 156167 & 4XMMJ171548.7+044742 & 258.953 & 4.795 & F7V & 6092 & 85 & 0724380101 & 29 & 1688 \\ \hline
HD 157720 & 4XMMJ172637.5-381007 & 261.656 & -38.169 & F5V & 6462 & 101 & 0724220101 & 28 & 498 \\ \hline
HD 159981 & 4XMMJ173859.1-342903 & 264.747 & -34.484 & F7V & 6321 & 72 & 0803030201 & 22 & 2797 \\ \hline
HD 193825 & 4XMMJ202303.3-205044  & 305.764 & -20.846 & F7V & 6042 & 207 & 0201902301 & 21 & 717 \\ \hline
HD 199544 & 4XMMJ205919.0-425311 & 314.829 & -42.886 & F8/G0V & 6067 & 136 & 0691670101 & 50 & 680 \\ \hline 
HD 219080 & 4XMMJ231233.2+492424 & 348.139 & 49.407 & F1V & 7069 & 25 & 0801010501 & 20 & 658 \\ \hline
HD 219570 & 4XMMJ231717.7-545604 & 349.324 & -54.935 & F6V & 6204 & 78 & 0505382201 & 16 & 469 \\ \hline
HD 222545 & 4XMMJ234135.8-084931 & 355.399 & -8.825 & F3V & 6478 & 113 & 0693010801 & 17 & 553 \\ \hline

HD 217566 & 4XMMJ230158.3-394521 &  345.493 & -39.756 & F3V & 6509 & 131 & 0824450401 & 77 & 1460 \\ \hline

\end{tabular}
\label{tab:obssummary}
\begin{flushleft} 
\footnotesize{
\hspace{0.5cm}$^a$ Target name of the corresponding X-ray source tabulated in \citet{2020A&A...641A.136W}. \\
\hspace{0.5cm}$^b$ Spectral type based on the SIMBAD astronomical database. \\
\hspace{0.5cm}$^c$ Effective temperature and distance based on \citet{2016A&A...595A...1G,2021yCat.1352....0B}.\\
\hspace{0.5cm}$^d$ The {\it XMM-Newton} observation identification. \\
\hspace{0.5cm}$^e$ Net exposure time of the PN detector used in our spectral analysis. \\}
\end{flushleft}
  \end{center}
\end{table*}

\section{Analysis and results} \label{sec:analysis_results}
In this section, to investigate the X-ray spectroscopic properties of single F-type MS stars, 
we performed spectral fitting for each sample and showed their results.
The spectral fitting procedure performed here is similar to that in our previous work \citet{2020ApJ...901...70T}.
First, we created a redistribution matrix and ancillary response files 
which contain the detector response, the correction matrix file for the incident X-ray energy, 
and the effective area of the telescope for each detector, respectively. 
Then, the spectra were extracted from a circle with a radius of 30 arcseconds centered on the coordinates listed in the X-ray source catalog. 
The background spectrum was extracted from an annular region between 45 and 60 arcseconds in radius 
centered on the same coordinates as those of the X-ray sources to consider the position dependence of the detector background. 
Next, the background spectrum was subtracted from the spectrum of the X-ray source. 
We confirmed that no bright point sources exist and there are no diffuse sources even in the background area 
as described in \S \ref{sec:obs_datareduction}.

\begin{figure*}[h!]
\begin{tabular}{ccc}
\begin{minipage}{0.45\hsize}
\hspace{0.5cm}
(a) HD222545 (Obs. ID:0693010801)
\vspace{-0.3cm}
\begin{center}
    \includegraphics[width=0.85\linewidth]{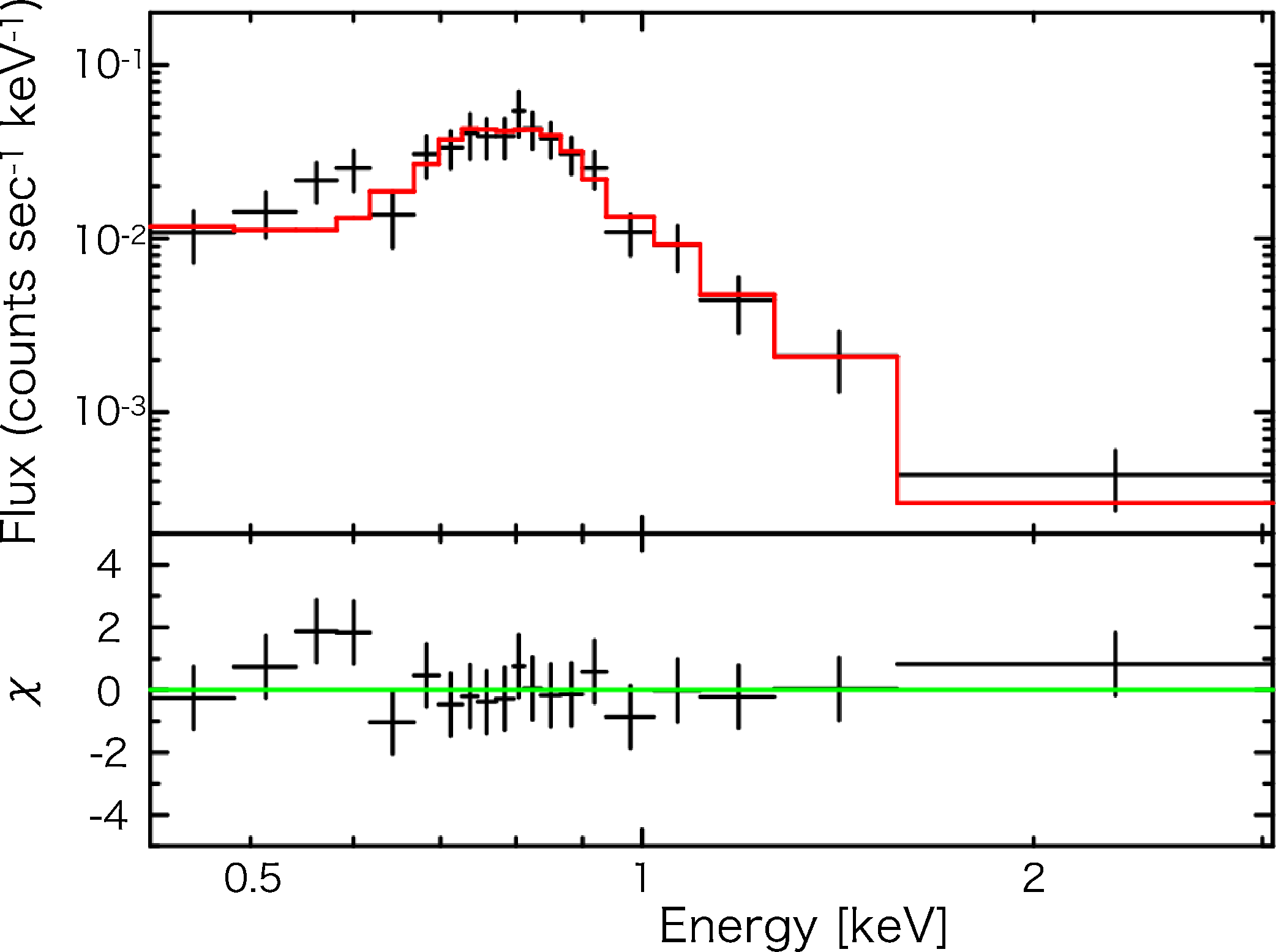}
\end{center}
\end{minipage}
\begin{minipage}{0.45\hsize}
\hspace{0.5cm}
(b) HD14245 (Obs. ID:0760540101) 
\begin{center}
\vspace{-0.3cm}
    \includegraphics[width=0.85\linewidth]{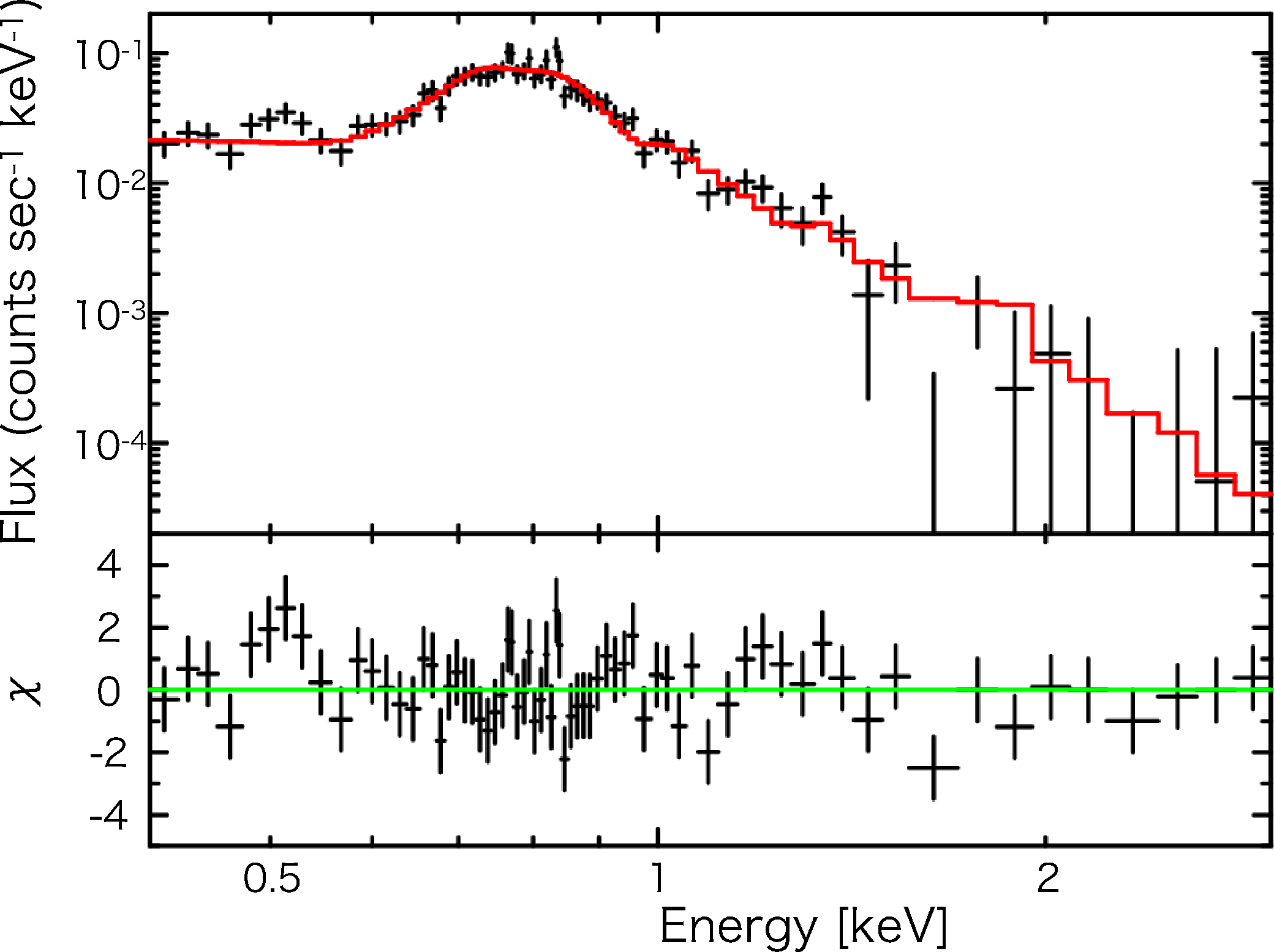}
\end{center}
\end{minipage}
\vspace{0.2cm}
\\
\begin{minipage}{0.45\hsize}
\hspace{0.5cm}
(c) HD159981 (Obs. ID:0803030201)
\begin{center}
\vspace{-0.3cm}
    \includegraphics[width=0.85\linewidth]{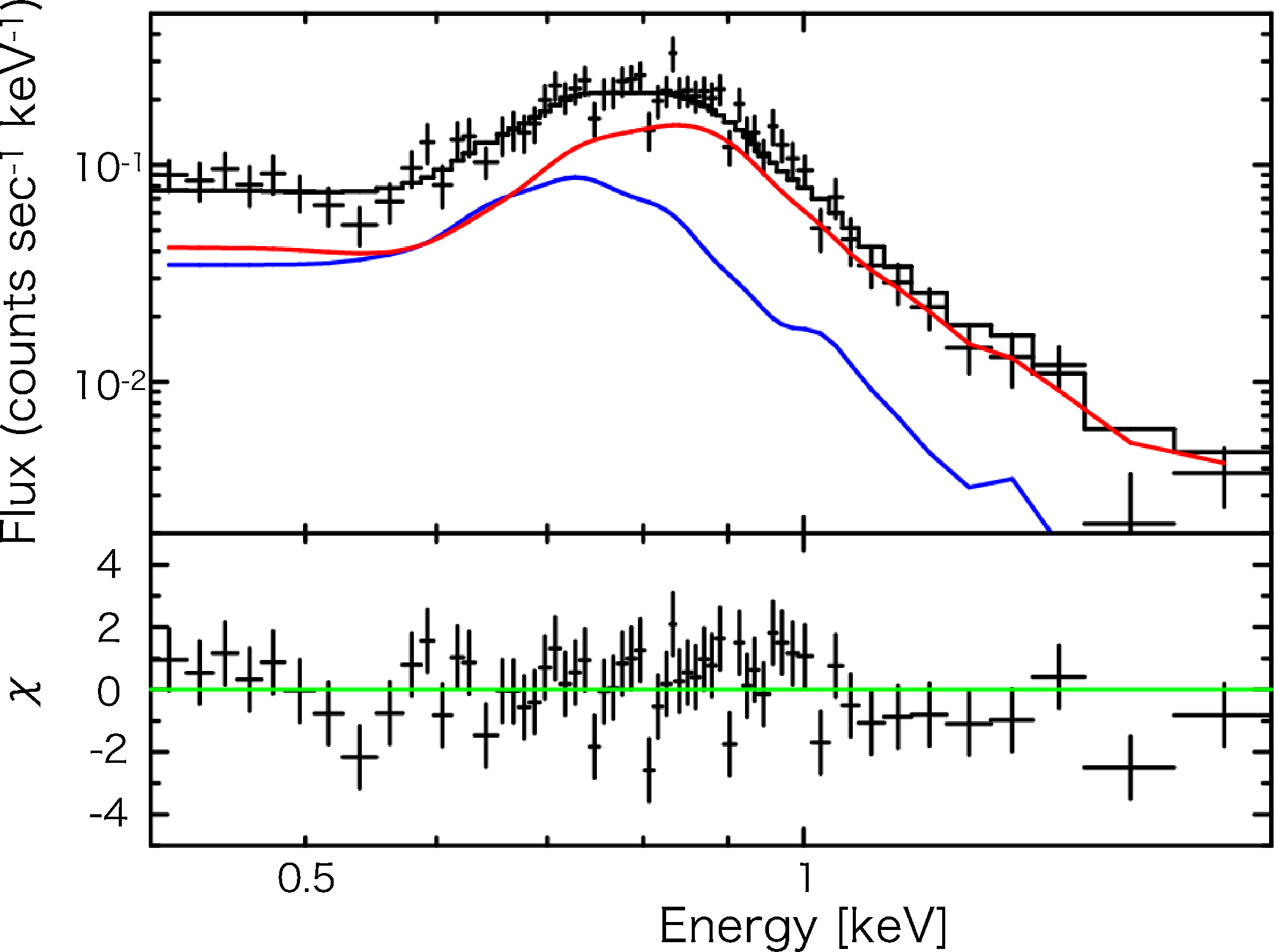}
\end{center}
\end{minipage}
\vspace{0.2cm}
\end{tabular}
  \caption{Examples of the spectra for the one-temperature model with the relatively poor (top left) and rich (top right) photon statistics and the two-temperature model (bottom). 
  The best fit models are also shown and the cooler and hotter components are indicated by the blue and red lines for the two-temperature model.}
  \label{fig:example_spectra}
\end{figure*}

As is the case with previous X-ray spectroscopic studies for stellar coronae \citep[see, e.g., literature in][]{2009A&ARv..17..309G,2020ApJ...901...70T}, 
we also used one-, two- or three-temperature, optically-thin thermal plasma models assuming the collisional ionization equilibrium state 
by using the astrophysical plasma emission code, APEC \citep{2001ApJ...556L..91S} with a $\chi$-square statistic.
In our spectral analysis, the interstellar hydrogen column density ($N_{\rm H}$) was estimated from the visual extinction derived from the three-dimensional dust map of \citet{Edenhofera&a2024}, using the empirical relation of \citet{10.1111/j.1365-2966.2009.15598.x}. 
For the two targets within 69 pc, outside the distance range covered by the dust map, we assumed $N_{\rm H}=10^{18.5}$ cm$^{-2}$. 
We confirmed that the estimated $N_{\rm H}$ values were consistent with those obtained from the spectral fits with $N_{\rm H}$ left free, within the statistical uncertainties. 
We therefore fixed $N_{\rm H}$ at the estimated value for each target.
In addition, following previous X-ray studies of FGK-type MS stars (e.g., \citealt{2026arXiv260727832B}), we fixed the coronal abundance at $Z=0.3Z_{\odot}$, using the solar photospheric elemental abundances of \citet{1989GeCoA..53..197A} as the reference. We confirmed that the resulting spectral parameters were consistent with those obtained when the abundance was allowed to vary, within the statistical uncertainties. We therefore adopted the fitting results with both $N_{\rm H}$ and $Z$ fixed as our final results.
%
%
These values are assumed to be the same among the plasmas in multiple temperature models. 
The two- or three-temperature plasma model was adopted only if the new model improves the fitting 
with a significance level of $\gtrsim$99\%.
Considering our scientific goals through the high-quality spectral analysis, 
only the stars with small error bars (relatively within a factor of 2 in temperature and two orders of magnitude in normalization between the minimum and maximum values) 
were included in our final sample.
In determining the X-ray luminosity from the derived X-ray flux,
we used the distance obtained from {\it Gaia} satellite observations \citep{2021AJ....161..147B}. 
The EM serves as an indicator of X-ray activity, representing the amount of thermal plasma integrated along the line of sight. Also, it is a useful diagnostic for constraining the frequency of heating events in solar and stellar coronal structures \citep[e.g.][]{Athiray_2024}.
Note the uncertainty in the EM and X-ray luminosity due to the error in the distance is not considered in our results because the error is mostly less than 1\%, 
which is much smaller than the statistical error for the luminosity in our spectral analysis. 
Figure \ref{fig:example_spectra} shows three examples of the observed spectra 
with the best fit models for the one-temperature model with relatively poor (HD222545) and rich (HD14245) photon statistics and the two-temperature model (HD159981).
The parameters derived from the spectroscopic analysis are summarized in Table \ref{table:results}.
The adopted interstellar hydrogen column densities and stellar radii used to calculate the column EM are also listed in Table \ref{table:results}.
The one- and two-temperature plasmas were adopted for 32 and 1 targets, respectively 
while the three-temperature model was not used. 
For the one-temperature model, the plasma temperature ranged from 0.3 to 0.9 keV 
in their best-fit values
.
The observed unabsorbed total X-ray luminosity ranges from 9 $\times$ 10$^{27}$ to 8 $\times$ 10$^{29}$ erg~s$^{-1}$, 
while the emission measure from 8 $\times$ 10$^{50}$ to 5 $\times$ 10$^{52}$ cm$^{-3}$ in their best-fit values.
For the two-temperature model, 
the temperature and EM values are 0.3 keV and 4 $\times 10^{51}$ cm$^{-3}$ for the low-temperature plasma and 
0.8 keV and 4 $\times 10^{51}$ cm$^{-3}$ for the high-temperature plasma, respectively , while the total X-ray luminosity is $1.5 \times 10^{29}$ erg s$^{-1}$.
In this case, we adopted the EM-weighted temperature in the same manner as \citet{Johnstone2015} 
and the weighted temperature is 0.5 keV.
As reported earlier \citep{gudel1997,2020ApJ...901...70T}, 
we evaluated the systematic errors on the abundance table and the plasma emission code used above.
An alternative abundance table \citep{asplund2009} and plasma emission code, MEKAL \citep{mekal} were used, and we confirmed that 
resultant parameters are consistent with each other between before and after the replacement within the statistical errors 
except for the temperature between the two codes.
The temperature obtained from the mekal model is systematically 10--20\% lower than that obtained from the apec model and 
the same trend is reported in previous studies \citep{gudel1997,2020ApJ...901...70T}.
The difference does not affect our discussions and conclusions in the next section and thus the results with the apec model are used in our discussions.
Note that our spectral analysis results are consistent with previous studies within the statistical error \citep{1999A&A...348..161P,2000ApJS..131..335R,2012ApJ...756...27L} 
even though the detailed spectral analysis was not performed and there remains the possibility that the sample includes unresolved multiple systems, in which case the observed emission represents the sum of all components.
\begin{table*}[htb]
\scriptsize{
  \caption{The best fit parameters of the model fitting for F-type MS stars.
  The hydrogen column density was fixed at the adopted value for each target, and the metal abundance was fixed at $0.3Z_\odot$.}
\label{table:results}
  \begin{center}
    \setlength{\tabcolsep}{3pt} 
    \begin{tabular}{@{}lllllllll@{}}
\hline\hline
Target Name & Date & $R_\ast$ & $\log N_{\rm H}$$^a$ & $\bar{kT}$$^b$ & EM & $L_{\rm X}$ & $\chi^2/$d.o.f & $P_{\rm rot}$$^c$ \\
            & yyyy/mm & [$R_\odot$] & [cm$^{-2}$] & [keV] & [10$^{52}$cm$^{-3}$] & [10$^{29}$~erg~s$^{-1}$] & & [day] \\ \hline \hline
HD 224997 & 2010/06 & 1.7 & 19.9 & $0.64_{-0.08}^{+0.07}$ & $0.80_{-0.09}^{+0.09}$ & $1.3_{-0.1}^{+0.1}$ & 22/29 & N/A \\ \hline
HD 5677 & 2003/12 & 2.0 & 19.8 & $0.54_{-0.08}^{+0.07}$ & $0.48_{-0.06}^{+0.08}$ & $0.72_{-0.09}^{+0.09}$ & 38/25 & 1.3 \\ \hline
HD 7779 & 2016/11 & 1.6 & 20.1 & $0.66_{-0.08}^{+0.07}$ & $0.83_{-0.09}^{+0.09}$ & $1.3_{-0.2}^{+0.2}$ & 24/20 & N/A \\ \hline
HD 13456 & 2004/01 & 2.2 & 18.5 & $0.43_{-0.03}^{+0.04}$ & $0.92_{-0.07}^{+0.07}$ & $1.2_{-0.1}^{+0.1}$ & 28/34 & 1.2 \\ \hline
HD 14224$^d$ & 2016/07 & 1.7 & 19.8 & $0.63_{-0.07}^{+0.07}$ & $0.77_{-0.08}^{+0.08}$ & $1.2_{-0.1}^{+0.1}$ & 17/20 & 2.9-5.7 \\ \hline
HD 14245$^d$ & 2015/07 & 1.2 & 19.7 & $0.55_{-0.03}^{+0.03}$ & $0.92_{-0.04}^{+0.04}$ & $1.4_{-0.1}^{+0.1}$ & 82/69 & 2.2-2.9 \\ \hline
HD 15557 & 2013/02 & 2.4 & 19.8 & $0.62_{-0.08}^{+0.05}$ & $0.87_{-0.07}^{+0.07}$ & $1.4_{-0.1}^{+0.1}$ & 68/32 & N/A \\ \hline
HD 23352 & 2015/02 & 1.4 & 20.4 & $0.67_{-0.04}^{+0.04}$ & $1.8_{-0.1}^{+0.1}$ & $2.9_{-0.2}^{+0.2}$ & 41/44 & N/A \\ \hline
HD 32848$^d$ & 2003/09 & 1.3 & 19.6 & $0.63_{-0.03}^{+0.03}$ & $1.7_{-0.1}^{+0.1}$ & $2.8_{-0.1}^{+0.1}$ & 51/61 & 2.1-2.7 \\ \hline
HD 32645 & 2009/02 & 1.5 & 19.6 & $0.64_{-0.03}^{+0.03}$ & $2.8_{-0.1}^{+0.1}$ & $4.4_{-0.2}^{+0.2}$ & 50/54 & 0.2-0.5 \\ \hline
HD 36364 & 2016/03 & 3.2 & 20.3 & $0.76_{-0.03}^{+0.03}$ & $4.0_{-0.2}^{+0.2}$ & $6.6_{-0.4}^{+0.4}$ & 51/53 & N/A \\ \hline
HD 37552 & 2012/03 & 1.5 & 19.5 & $0.69_{-0.11}^{+0.09}$ & $0.64_{-0.10}^{+0.10}$ & $1.0_{-0.2}^{+0.2}$ & 32/26 & 2.1-5.8 \\ \hline
HD 37916 & 2015/03 & 1.5 & 19.7 & $0.36_{-0.03}^{+0.03}$ & $0.83_{-0.08}^{+0.08}$ & $1.0_{-0.1}^{+0.1}$ & 23/30 & 0.9-1.0 \\ \hline
HD 39904 & 2012/12 & 1.4 & 19.5 & $0.63_{-0.08}^{+0.07}$ & $0.56_{-0.05}^{+0.05}$ & $0.88_{-0.08}^{+0.08}$ & 34/29 & 0.8-1.0 \\ \hline
BD+19 2050 & 2015/05 & 1.6 & 19.8 & $0.41_{-0.05}^{+0.08}$ & $1.9_{-0.3}^{+0.3}$ & $2.4_{-0.2}^{+0.2}$ & 46/36 & N/A \\ \hline
BD+20 2176 & 2015/04 & 1.4 & 19.9 & $0.61_{-0.06}^{+0.05}$ & $1.2_{-0.1}^{+0.1}$ & $1.8_{-0.2}^{+0.2}$ & 38/33 & N/A \\ \hline
BD+20 2180 & 2015/04 & 1.4 & 19.9 & $0.61_{-0.05}^{+0.04}$ & $0.81_{-0.06}^{+0.07}$ & $1.3_{-0.1}^{+0.1}$ & 31/26 & N/A \\ \hline
HD 73937 & 2015/05 & 1.7 & 19.9 & $0.56_{-0.10}^{+0.06}$ & $1.5_{-0.1}^{+0.2}$ & $2.2_{-0.2}^{+0.2}$ & 55/33 & N/A \\ \hline
HD 104456 & 2002/01 & 1.8 & 19.7 & $0.63_{-0.08}^{+0.07}$ & $0.41_{-0.04}^{+0.04}$ & $0.65_{-0.06}^{+0.06}$ & 34/29 & N/A \\ \hline
HD 114746$^d$ & 2008/01 & 2.7 & 20.6 & $0.73_{-0.04}^{+0.04}$ & $1.5_{-0.1}^{+0.1}$ & $2.5_{-0.2}^{+0.2}$ & 40/44 & 3.6 \\ \hline
HD 117822$^d$ & 2003/01 & 1.6 & 20.1 & $0.68_{-0.10}^{+0.09}$ & $0.54_{-0.07}^{+0.07}$ & $0.87_{-0.12}^{+0.12}$ & 28/32 & 1.1 \\ \hline
HD 119718$^d$ & 2014/07 & 2.2 & 20.0 & $0.62_{-0.08}^{+0.06}$ & $1.3_{-0.1}^{+0.1}$ & $2.0_{-0.2}^{+0.2}$ & 31/26 & 1.1 \\ \hline
HD 122390 & 2010/01 & 1.4 & 20.3 & $0.62_{-0.03}^{+0.03}$ & $2.0_{-0.1}^{+0.1}$ & $3.2_{-0.1}^{+0.1}$ & 72/66 & N/A \\ \hline
HD 150345 & 2001/09 & 1.4 & 19.5 & $0.88_{-0.03}^{+0.03}$ & $2.9_{-0.2}^{+0.2}$ & $4.6_{-0.3}^{+0.3}$ & 39/38 & N/A \\ \hline
HD 156167 & 2013/09 & 1.2 & 19.7 & $0.63_{-0.03}^{+0.03}$ & $2.1_{-0.1}^{+0.1}$ & $3.3_{-0.1}^{+0.1}$ & 72/49 & N/A \\ \hline
HD 157720$^d$ & 2013/09 & 1.4 & 19.5 & $0.58_{-0.07}^{+0.06}$ & $0.73_{-0.07}^{+0.07}$ & $1.1_{-0.1}^{+0.1}$ & 24/38 & 1.8-1.9 \\ \hline
HD 159981$^{d,e}$ & 2018/03 & 1.8 & 19.3 & $0.57_{-0.04}^{+0.04}$ & $1.1_{-0.2}^{+0.2}$ & $1.6_{-0.3}^{+0.3}$ & 73/56 & 4.8 \\ \hline
HD 193825 & 2005/04 & 2.2 & 20.5 & $0.72_{-0.04}^{+0.05}$ & $4.8_{-0.3}^{+0.3}$ & $7.8_{-0.5}^{+0.5}$ & 41/42 & N/A \\ \hline
HD 199544$^d$ & 2013/04 & 1.3 & 19.7 & $0.65_{-0.05}^{+0.05}$ & $0.80_{-0.07}^{+0.07}$ & $1.3_{-0.1}^{+0.1}$ & 33/35 & 3.9-6.2 \\ \hline
HD 219080 & 2017/12 & 1.9 & 18.5 & $0.25_{-0.01}^{+0.02}$ & $0.081_{-0.007}^{+0.007}$ & $0.091_{-0.008}^{+0.008}$ & 34/29 & 0.5 \\ \hline
HD 219570$^d$ & 2007/10 & 1.3 & 19.6 & $0.56_{-0.10}^{+0.08}$ & $0.47_{-0.05}^{+0.06}$ & $0.72_{-0.06}^{+0.06}$ & 23/31 & 6.2 \\ \hline
HD 222545$^d$ & 2012/12 & 1.6 & 19.7 & $0.61_{-0.07}^{+0.06}$ & $0.97_{-0.10}^{+0.11}$ & $1.5_{-0.2}^{+0.2}$ & 12/18 & 1.5-2.1 \\ \hline
HD 217566$^d$ & 2018/05 & 2.3 & 19.6 & $0.62_{-0.04}^{+0.03}$ & $0.91_{-0.05}^{+0.05}$ & $1.4_{-0.1}^{+0.1}$ & 104/84 & 2.0-3.4 \\ \hline
\hline
    \end{tabular}
  \end{center}}
\begin{flushleft}
\footnotesize{
\hspace{3.5cm}$^a$ $\log N_{\rm H}$ denotes $\log_{10}(N_{\rm H}/{\rm cm}^{-2})$. The hydrogen column densities were estimated from the visual extinction obtained using the three-dimensional dust map of \citet{Edenhofera&a2024}, converted to $N_{\rm H}$ using the empirical relation of \citet{10.1111/j.1365-2966.2009.15598.x}. For targets closer than 69 pc, $\log N_{\rm H}=18.5$ was adopted. \\
\hspace{3.5cm}$^b$ EM-weighted temperature. \\
\hspace{3.5cm}$^c$ Rotation period estimated from the {\it TESS} satellite \citep{2014SPIE.9143E..20R} and all the {\it TESS} data used in this paper can be found in MAST: \dataset[10.17909/hvwa-m790]{http://dx.doi.org/10.17909/hvwa-m790}. \\
\hspace{3.5cm}$^d$ Target plotted in Figure \ref{fig:rossbynumber} with relatively small error bars on the Rossby number calculated in the discussion section. \\
\hspace{3.5cm}$^e$ Two-temperature model was adopted and the EM-weighted temperature is calculated in the same manner as \citet{Johnstone2015}. \\
}
\end{flushleft}
\end{table*}

\section{Discussion} \label{sec:discussion}
In this section, we compare our spectral analysis results to previous observational studies for the single G-dwarf stars 
and the theoretical models by extending original models established for G-type MS stars to F-type MS stars to discuss the spectral features and the X-ray coronal properties.
Additionally, the Rossby number for F-type MS stars is also calculated as supplemental data for our discussions to compare with those of late-type stars.

\begin{figure*}
\epsscale{1.0}
\plotone{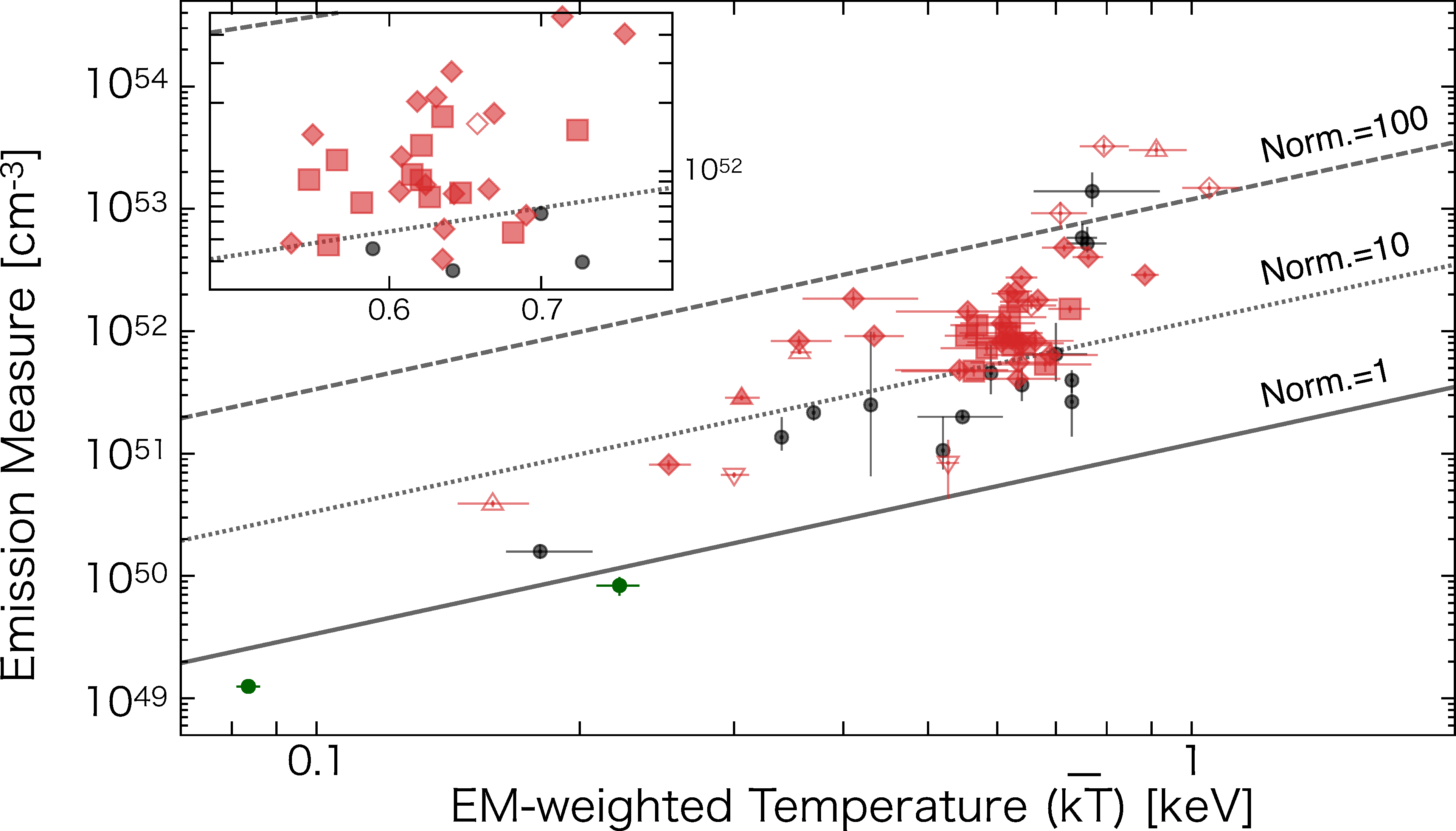}
\plotone{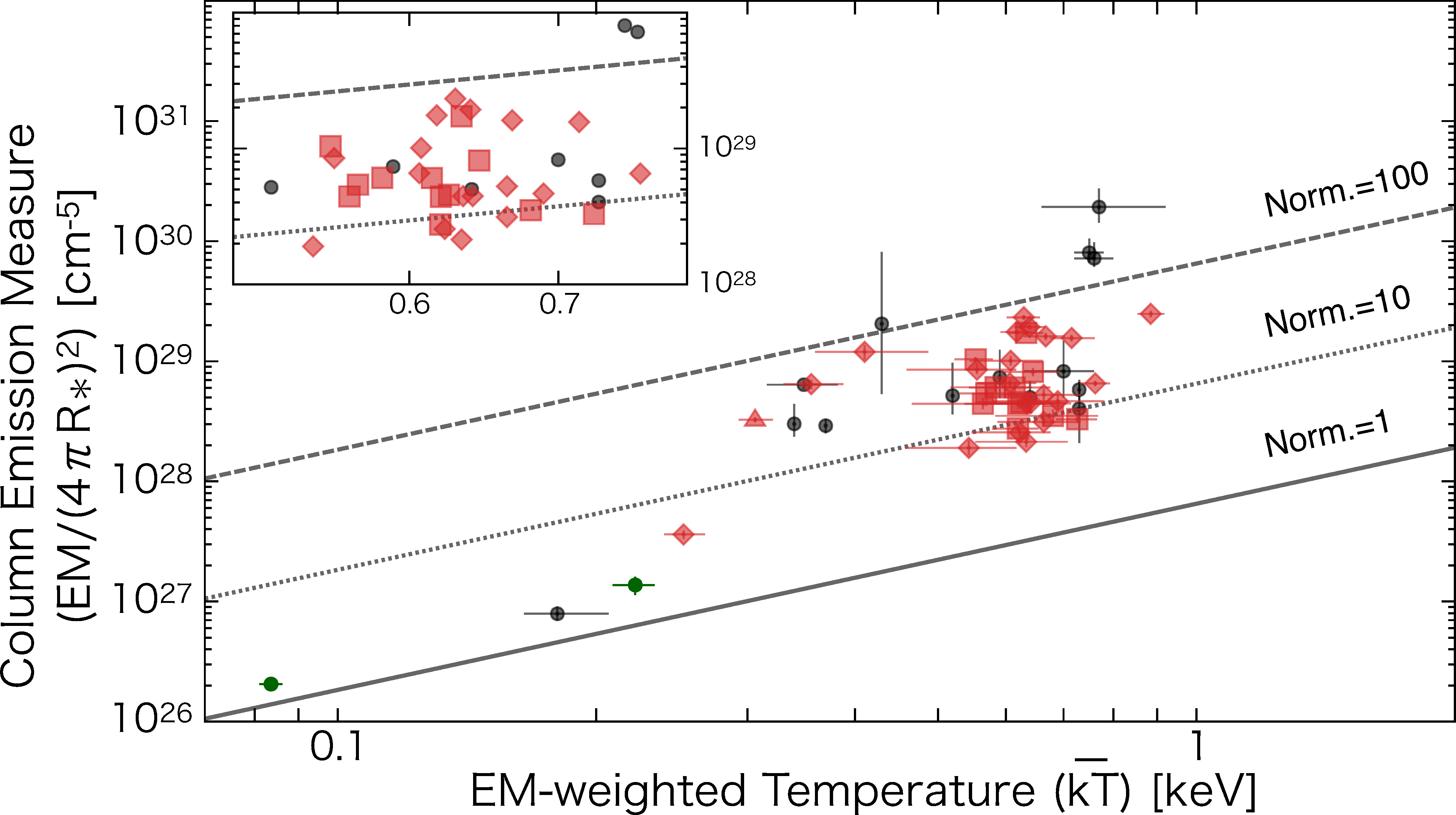}
\caption{Resultant (a) EM--$\bar{kT}$ and (b) column EM~(${\rm EM}/4\pi R_*^2$)--$\bar{kT}$ relations for single (filled) and binary (open) F-type MS (red) stars with a close-up view of the complex area without error bars for visual simplicity.
Binary stars are included only in panel (a), whereas panel (b) shows single stars only. Here, $R_*$ denotes the radius of each
individual star.
The stars obtained through the cross-matching between the {\it XMM-Newton} source catalog and the Tycho-2 spectral type catalog are shown as diamonds, however, the stars also plotted in Figure \ref{fig:rossbynumber} are displayed as squares instead of diamonds.
The stars obtained through the cross-matching between the {\it XMM-Newton} source catalog and the GLIESE2MAS catalog are shown as triangles and analyzed in previous studies and this work are
displayed as up- and down-pointing triangles, respectively.
The solid, dotted, and dashed lines are our theoretical models with different normalization values of 1, 10, and 100, respectively, calculated assuming a representative stellar radius of $1.7R_\odot$, the mean radius of our 33 single F-type MS stars. In panel (b), the theoretical EM is divided
by the surface area, $4\pi(1.7R_\odot)^2$, whereas the observed EM is divided by $4\pi R_*^2$ for each star.
Here we adopt $f=0.1$, $\bar{B}=100~{\rm G}$ and $F_{\rm h}=10^7~{\rm erg~cm^{-2}~s^{-1}}$ as typical values of a filling factor, a magnetic field strength, and a heating flux of the active regions, respectively.
For comparison, spectral analysis results for single G dwarf stars are shown in black \citep{gudel1997,2020ApJ...901...70T}, while the Sun is shown in dark green, with the solar cycle maximum and minimum explicitly labeled \citep{Peres2000}. 
}
\label{fig:em-kt}
\end{figure*}

To discuss the spectral features, we focused on the EM--$\bar{kT}$ and column EM--$\bar{kT}$ relations, where the column EM is defined as ${\rm EM}/(4\pi R_*^2)$ using the radius of each star, and Figure \ref{fig:em-kt} shows these relationships for our targets. 
Binary systems are included in the EM--$\bar{kT}$ relation, whereas the column EM--$\bar{kT}$ relation is restricted to single stars because of the ambiguity in assigning stellar radii to the X-ray-emitting components of binary systems.
The red-filled marks represent the single F-type MS stars in this study.
As is the case with the $L_{\rm X}$--$\bar{kT}$ relation, there is a clear positive correlation as seen in \citet{gudel1997,Johnstone2015,2020ApJ...901...70T} 
for G dwarf stars and late-type stars and the correlation coefficient is 0.50.
We also compared our theoretical models (originally constructed for single G-type MS stars based on theoretical and observational studies for the Sun)
to the relationships obtained for the single F-type MS stars.
For the theoretical curves in both the EM--$\bar{kT}$ and column EM--$\bar{kT}$ relations, we adopted a representative stellar radius of $1.7R_\odot$, corresponding to the mean radius of our 33 single F-type MS stars. For the column EM relation, the theoretical EM was divided by $4\pi(1.7R_\odot)^2$, whereas the observed EM was divided by the surface area of each individual star, $4\pi R_*^2$.
The other parameters, such as the filling factor, the heating flux, and the magnetic field strength of the active regions were assumed to be typical values of G-type MS stars ($f=0.1$, $\bar{B}=100~{\rm G}$ and $F_{\rm h}=10^7~{\rm erg~cm^{-2}~s^{-1}}$).
%
The power-law index for the size distribution function was also assumed to be the same as the observed solar value.
%
%
Resultant theoretical lines with different normalization values of 1, 10, and 100 are plotted in Figure \ref{fig:em-kt}.
The normalization determines the total number of active regions and the normalization of 1 corresponds to the case of the Sun.
The theoretical curves in the EM--$\bar{kT}$ relation are calculated for the solar case, while those in the column EM--$\bar{kT}$ relation are normalized by the stellar surface area.
The details are described in \citet{2020ApJ...901...70T}.
%
%
Consequently, the following three observational features were found:
(1) our single F-type MS stars have a plasma with an EM-weighted temperature of $\lesssim$1 keV, 
(2) an EM of $\lesssim$10$^{53}$ cm$^{-3}$ corresponding to $L_{\rm X}$ of $\lesssim$10$^{30}$~erg~s$^{-1}$, and (3) the stars are included in the range between normalizations of roughly 10 and 100 
in our theoretical curves.
The last characteristic suggests that the stars in this study are much brighter than the ones expected from the Sun.
For comparison, X-ray spectral characteristics for the single G dwarf stars in \citet{2020ApJ...901...70T} 
adopting mostly the same catalogs and sample selection criteria are also plotted and 
these observational aspects of (1), (2), and (3) are consistent with those of the single G dwarf stars.
Thus, these observational analogies suggest that the overall X-ray coronal properties of single G dwarf and F-type MS stars are broadly similar.
The column EM--$\bar{kT}$ relation also shows broadly similar distributions for single F- and G-type stars. However, some single G-type stars exhibit higher column EM values than those of the single F-type stars in our sample. This difference may reflect variations in the surface coverage and physical
properties of coronal active regions, which are closely related to the surface
X-ray activity \citep{Johnstone2015,Woodaj2018}. Such variations may also be associated with differences in magnetic dynamo properties between F- and G-type stars, possibly related to their different convective properties \citep{2011ApJ...741...54C}.
Spectral analysis results for the single G-type MS stars in a previous study \citep{gudel1997} including the solar maximum and minimum periods \citep{Peres2000}
are also shown in Figure \ref{fig:em-kt}. 
Note that only the single stars meeting our criteria for selection were plotted for comparison. 

To verify our statements above and discuss whether the observational features apply exclusively to single stars, two approaches were attempted to search for additional F-type MS star systems 
by extending our sample selection conditions and using a different star catalog. 
To this end, first, we picked up the brightest stars with an expected X-ray luminosity of $\gtrsim$10$^{30}$~erg~s$^{-1}$ 
based on the observed flux in the X-ray source catalog without any classifications of single and binary star systems using the same catalogs.
In total 5 objects were picked up and all of them were classified as binary systems and have rich photon statistics. 
These results are consistent with those of \citet{1998A&AS..132..155H}, which provided a catalog of optically bright MS stars detected by {\it ROSAT} 
including the information of their X-ray luminosities estimated from count rates and hardness ratios without the detailed spectral analysis.
In the catalog of \citet{1998A&AS..132..155H}, all the F-type stars with a luminosity of $\gtrsim$10$^{30}$~erg~s$^{-1}$ satisfy our criteria for binary systems.
Note that, owing to poor photon statistics, no fainter sample was added.
Second, we selected a star catalog after the cross identifications 
between the CNS3 (Gliese Catalog of Nearby Stars, 3rd Edition) \citep{1991adc..rept.....G} and the 2MASS (Two Micron All Sky Survey) \citep{2006AJ....131.1163S} catalogs \citep{2010PASP..122..885S} 
because the catalog (GLIESE2MAS) contains only nearby stars within 25 parsecs and thus brighter and fainter stars are expected to be observed.
After the same sample selection procedure was conducted, 
6 objects including 1 single and 5 binary star systems were selected.
For the 11 additional objects, spectral analysis was performed for 6 objects by ourselves, while the spectral parameters for the other 5 objects were adopted from previous studies: HD 61421 (Procyon; \citet{2002A&A...389..228R}), HD 120136 ($\tau$ Boo; \citet{2011A&A...532A...6S}), HD 146361 ($\sigma$ CrB A; \citet{2005ApJ...630.1074S}), HD 17051 ($\iota$ Hor; \citet{2019A&A...631A..45S}), and HD 179949 \citep{2013A&A...552A...7S}. All 11 objects are shown in Figure \ref{fig:em-kt}, while Table \ref{table:add_results} summarizes only the 6 objects for which we performed the spectral analysis in this study, including their adopted distances and stellar radii.
For HD 97975 and HD 152041, the spectral fits were not adequately reproduced even when the metal abundances were fixed at their Gaia-based photospheric values. We therefore allowed the abundances to vary and obtained
$Z=0.15\pm0.02Z_\odot$ and $Z=0.13^{+0.05}_{-0.03}Z_\odot$, respectively.
%
%


\begin{table*}[htb]
\scriptsize{
  \caption{The best fit parameters of the model fitting conducted in our study for the additional F-type MS stars.
  The hydrogen column density was fixed for all targets, while the metal abundance was fixed at $0.3Z_\odot$ except for HD 97975 and HD 152041, for which it was allowed to vary.}
\label{table:add_results}
  \begin{center}
    \setlength{\tabcolsep}{3pt}
    \begin{tabular}{@{}llllllllll@{}}
\hline\hline
Target Name & Date & dist. & $R_\ast$ & $\log N_{\rm H}$$^a$ & $\bar{kT}$ & EM & $L_{\rm X}$ & $\chi^2/$d.o.f & $P_{\rm rot}$$^b$ \\
            & yyyy/mm & [pc] & [$R_\odot$] & [cm$^{-2}$] & [keV] & [10$^{52}$cm$^{-3}$] & [10$^{29}$~erg~s$^{-1}$] & & [day] \\ \hline\hline
\multicolumn{10}{c}{{\it XMM-Newton} source catalog~$\times$~Tycho-2 spectral type catalog (brightest stars)} \\ \hline
CD-52 2501$^{c}$ & 2001/11 & 151 & 1.3 & 19.8 & $0.71_{-0.05}^{+0.05}$ & $9.2_{-2.3}^{+2.2}$ & $14_{-4}^{+7}$ & 62/55 & N/A \\ \hline
BD+20 2189 & 2015/05 & 162 & 1.5 & 19.9 & $0.66_{-0.05}^{+0.04}$ & $1.6_{-0.1}^{+0.1}$ & $2.6_{-0.2}^{+0.2}$ & 43/35 & N/A \\ \hline
HD 86500$^{c}$ & 2004/06 & 227 & 2.3 & 20.4 & $1.0_{-0.1}^{+0.1}$ & $15_{-2}^{+2}$ & $19_{-2}^{+2}$ & 55/47 & N/A \\ \hline
HD 97975$^{c}$ & 2012/06 & 227 & 2.7 & 20.1 & $0.79_{-0.05}^{+0.05}$ & $33_{-4}^{+5}$ & $35_{-5}^{+5}$ & 127/122 & N/A \\ \hline
HD 152041 & 2018/03 & 160 & 2.1 & 20.1 & $0.96_{-0.05}^{+0.06}$ & $79_{-11}^{+11}$ & $86_{-6}^{+6}$ & 26/17 & 0.8-0.9 \\ \hline
\hline
\multicolumn{10}{c}{{\it XMM-Newton} source catalog~$\times$~GLIESE2MAS catalog} \\ \hline
HD 9826 & 2013/08 & 13 & 1.7 & 18.5 & $0.30_{-0.01}^{+0.01}$ & $0.067_{-0.004}^{+0.004}$ & $0.076_{-0.004}^{+0.004}$ & 85/47 & N/A \\ \hline
\hline
    \end{tabular}
  \end{center}
}
\begin{flushleft}
\footnotesize{
\hspace{3.0cm}$^a$ $\log N_{\rm H}$ denotes $\log_{10}(N_{\rm H}/{\rm cm}^{-2})$. The hydrogen column densities were estimated from the visual extinction obtained using the three-dimensional dust map of \citet{Edenhofera&a2024}, converted to $N_{\rm H}$ using the empirical relation of \citet{10.1111/j.1365-2966.2009.15598.x}. For targets closer than 69 pc, $\log N_{\rm H}=18.5$ was adopted. \\
\hspace{3.0cm}$^b$ Rotation period estimated from the {\it TESS} satellite \citep{2014SPIE.9143E..20R} and all the {\it TESS} data used in this paper can be found in MAST: \dataset[10.17909/hvwa-m790]{http://dx.doi.org/10.17909/hvwa-m790}. \\
\hspace{3.0cm}$^c$ Two-temperature model was adopted and the EM-weighted temperature is calculated \\
\hspace{3.25cm}in the same manner as \citet{Johnstone2015}. \\
}
\end{flushleft}
\end{table*}

Consequently, some binary stars (possibly e.g., interactive or close binaries) have plasmas with a relatively higher temperature of up to nearly 1 keV, EMs of $\gtrsim$10$^{53}$~cm$^{-3}$ and $L_{\rm X}$ of $\gtrsim$10$^{30}$~erg~s$^{-1}$, and a normalization of $\gtrsim$100 
as previously reported for binary systems in different spectral types of stars \citep[e.g.][]{gudel1997}.
Moreover, no stars have the normalization values of $\ll$10.
This is caused by poor photon statistics and selection bias.
For example, the average distance of 140 pc and the minimum flux of 2$\times$10$^{-14}$~cm$^{-2}$~s$^{-1}$ in our sample 
correspond to a lower limit of an EM of 3$\times$10$^{51}$~cm$^{-3}$ assuming a typical abundance of 0.3 Z$_{\odot}$ and a temperature of 0.6 keV.
This means that, when using the combination of these stars (the Tycho-2 spectral type catalog) and X-ray source catalogs, it is difficult to detect fainter stars with a normalization value of $\ll$10.
Meanwhile, the combination of the nearby star (the GLIESE2MAS catalog) and the X-ray source catalogs is expected to detect much fainter stars 
because the distance is much smaller even though the number of registered stars is also smaller.
However, the averaged X-ray flux of the stars identified through the cross-match with the GLIESE2MAS catalog is 1$\times$10$^{-12}$~cm$^{-2}$~s$^{-1}$ 
because almost all of these stars were not detected serendipitously but were likely the main targets of the {\it XMM-Newton} observations, having already been identified as X-ray-bright sources through previous X-ray observations. 
This may indicate that the stars included in our study are preferentially selected from X-ray flux-bright objects.
In summary, feature (2), namely the upper limit of the EM ($\lesssim 10^{53}$ cm$^{-3}$ corresponding to $L_{\rm X}\lesssim10^{30}$ erg s$^{-1}$), appears to be a unique signature of single stars. Additionally, for feature (3), namely the distribution between normalization values of roughly 10 and 100, the upper bound is characteristic of single stars, whereas the lower bound is likely affected by the selection effect discussed above. 
The theoretical model of Takasao et al. (2020) predicts that (a) the upper limit of the EM-weighted temperature is determined by the size of the largest active region (see Equation (17) of \citet{2020ApJ...901...70T}), and (b) the total EM represents the sum of the contributions from individual active regions.
As for feature (1), a simple application of the coronal heating theory \citep{Rosner1978,Shibata2002} implies that an EM-weighted coronal temperature of $\sim$1~keV requires active regions with a size of $\sim$10$^{11}$ cm (see Equation (3) of \citet{2020ApJ...901...70T}), which is comparable to the stellar radius. Additionally, a temperature of $\sim$1~keV is similar to the virial temperature of F-type stars. These estimates suggest that the upper limit of the coronal temperature is determined by the stellar radius and/or gravitational binding. We note that the upper limit of the temperature may be set by nanoflare-heated active-region cores \citep{2011ApJ...740..111T,2012ApJ...759..141W} and/or unresolved stellar flares.
Similarly, the upper limit of the total EM is likely determined by the limited stellar surface area available for active regions. When the stellar surface is fully covered by active regions, the emergence of new active regions will be suppressed, which may potentially explain feature (2).

%
%
%
As discussed in \S \ref{sec:intro}, the convective turnover time is expected to be $\ll$1 day at approximately 6700 K.
Thus, we investigated the difference among our sample with T$_{\rm{eff}}$ $\geq$ 6700 K and T$_{\rm{eff}}$ $<$ 6700 K and 
found that late single F-type MS stars seem to distribute in a wide range of the temperatures.
However, because the sample is very limited and the uncertainty in the effective temperature is large for this discussion, 
we could not obtain any conclusive results.
A dedicated study focusing on earlier F-type stars (e.g., F0--F4) with a larger sample will be important future work.

%

Additionally, we focused on the Rossby number, the ratio of the rotation period to the convective turnover time (Ro$=P_\mathrm{rot}/\tau_{\mathrm{c}}$), 
as an important factor to discuss the coronal properties of the F-type MS stars.
The Rossby number plays an important role in dynamo models 
which are thought to be a common scenario resulting in magnetic activity and finally X-ray emission in low-mass MS stars \citep[e.g.][]{2003A&A...397..147P,2011ApJ...743...48W,2015RSPTA.37340259T}.
To determine the Rossby number of our sample, the rotation periods were obtained from {\it TESS} archival data 
as reported by \citet{2020ApJ...901...70T}.
For multiple data for one target, the averaged value was adopted as the nominal value and the minimum and the maximum values were taken into account 
as the error bars.
The effective temperature in \citet{2016A&A...595A...1G} was used and the uncertainty of a central 68\% confidence interval was also taken into account.
The convective turnover time was derived from an empirical relation described in \citet{2011ApJ...741...54C} 
by using two-color photometric data tabulated in \citet{2003AJ....125..359W} and the effective temperature.
The bolometric luminosity, $L_{\rm bol}$, was estimated from the Tycho-2 $B_{\rm T}$ and $V_{\rm T}$ photometry \citep{2003AJ....125..359W} with a bolometric correction based on the effective temperature and the stellar distance. 
The stellar radius, $R_*$, was then estimated from $L_{\rm bol}$ and the effective temperature using the Stefan--Boltzmann law. 
For stars for which some of the required catalog values were unavailable, alternative catalog or literature values were adopted. 
For HD 61421 and HD 9826, photometric values from the Gliese catalog \citep{1995yCat.5070....0G} were adopted. 
For HD 61421, the distance and effective temperature were taken from \citet{2002A&A...389..228R} and \citet{2004A&A...413..251K}, respectively. 
For CD-52 2501, the $B-V$ color was adopted from Gaia DR3 Performance Verification \citep{2022yCat.1360....0G}. 
%
%
For HD 2151, the stellar radius was directly adopted from \citet{10.1111/j.1745-3933.2007.00355.x}.

%

%
We discarded some cases with poor statistics because we could not constrain their Rossby numbers due to large uncertainties in the effective temperatures. Finally, we ended up with 16 objects, consisting of 12 single and 4 additional F-type MS stars.

\begin{figure*}
\epsscale{0.8}
\plotone{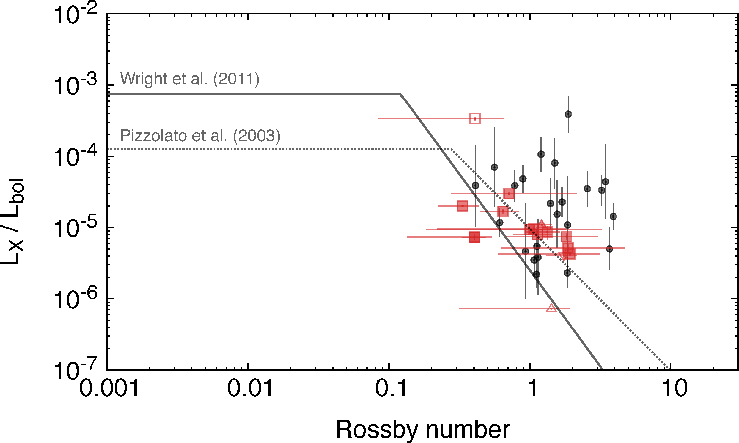}
\caption{
%
Resultant X-ray activity vs. Rossby number for F-type MS stars in this study (red). The symbols for the stars in this study follow the same convention as in Figure \ref{fig:em-kt}. The solid and dotted lines are the best-fit relations in previous studies for late-type stars \citep{2011ApJ...743...48W} and stars with masses of 1.10--1.29 $M_\odot$ \citep{2003A&A...397..147P}, respectively.
%
For comparison, data points obtained in a previous study \citep{2019A&A...628A..41P} for F-type MS stars are also plotted in black.}
\label{fig:rossbynumber}
\end{figure*}

The results are shown in Figure \ref{fig:rossbynumber}.
The solid line corresponds to the empirical best-fitting saturated and non-saturated activity--rotation relation observed for solar and late-type stars \citep{2011ApJ...743...48W}. The dotted line corresponds to the relation for approximately F-type MS stars (1.10--1.29 $M_\odot$) reported by \citet{2003A&A...397..147P}.
%
In both relations, all of our single F-type MS stars are in a non-saturated regime 
and the correlation coefficients of all of the single and both single and binary stars in our stars are $-0.48$ and $-0.36$ 
showing a negative correlation and the slope is consistent with those of both relations within statistical errors even though the error bars are very large.
Thus, it is found that there are no significantly different trends, 
suggesting that the X-ray coronal properties of the single F-type MS stars are not significantly different from those of late-type stars.
The distribution reported by \citet{2019A&A...628A..41P} for F-type main-sequence stars is also shown in Figure \ref{fig:rossbynumber} and appears to differ from our results. 
These differences may arise from variations in sample selection, such as the inclusion of binary systems and differences in flux and distance distributions. 
For instance, our study demonstrates F-type main-sequence stars have the upper limit of the EM of $\sim$10$^{53}$~cm$^{-3}$, corresponding to $L_{\rm X}$ of $\sim$10$^{30}$~erg~s$^{-1}$, suggesting that quiescent single F-type main-sequence stars cannot attain such large values of X-ray activity.
\citet{2019A&A...628A..41P} and this study derive the Rossby number in a similar manner, using the same empirical relation for $\tau_{\rm c}$ (T$_{\rm{eff}}$) \citep{2011ApJ...741...54C} and photometric light-curve analysis for $P_{\rm rot}$. Therefore, differences in the Rossby-number determination are unlikely to be the primary cause of the discrepancy.
In addition, differences in analysis methods, i.e., our use of spectroscopic analysis, also contribute to the discrepancy.
Note that, owing to the limited sample size, we were unable to examine differences between early (T$_{\rm{eff}}$ $\geq$ 6700\,K) and late (T$_{\rm{eff}}$ $<$ 6700\,K) stars 
and to compare the slope and normalization values in the unsaturated regime between this and previous studies.

Stellar winds, high-energy particles, and high-energy radiation such as X-rays and ultraviolet radiation can affect the physical and chemical structures of planetary atmospheres \citep[e.g.,][]{Airapetian+16, Linsky19, Owen19}. In particular, X-ray and UV irradiation can drive atmospheric escape (photoevaporation) of close-in planets \citep[e.g., ][]{Lammer+03,Owen+Jackson12,Kurokawa+Nakamoto14} \citep[see alternative processes in][]{Tanaka+14,Tanaka+15}. Since the X-ray absorption cross section and thus atmospheric heating depend on wavelength \citep[see discussions in][]{Owen+Alvarez16}, not only the X-ray luminosity but also the coronal temperature is important for evaluating the effects of stellar X-ray irradiation on planetary atmospheres.
%
The properties of such radiation from F-type stars have been poorly constrained, 
introducing uncertainties in escape estimates for their planets. 
Our study shows that F-type main-sequence stars exhibit X-ray properties similar to those of well-studied G-type stars, 
suggesting that atmospheric escape and planetary evolution around F-type stars can be modeled using frameworks developed for G-type stars. 
Recent work has shown that high-resolution MHD simulations can successfully reproduce the quasi-steady component of the X-ray and extreme ultraviolet (XUV) radiation spectra of solar-type stars \citep{2021A&A...656A.111S,2024A&A...691A.152S}. 
Our observational results therefore suggest that MHD models for solar-type stars can be applied to F-type main-sequence stars with appropriate parameter adjustments. 
An alternative empirical approach is to estimate the XUV flux by scaling the observed XUV spectra of representative F-type stars according to stellar radius and activity level. Recent examples of such stars are presented by \citet{2025A&A...693A.285S}, and this empirical approach will provide a useful comparison with our method in future work.

%
%
%

\section{Summary and Conclusions}\label{sec:summary-conclusions}
It is commonly recognized that stellar activity in the form of coronae plays a key role in a planetary system.
X-ray observations are powerful tools to investigate hot stellar plasmas, and X-ray spectral analysis allows us to quantitatively study and estimate their X-ray coronal properties and the impact on surroundings.
In this paper, we focused on X-ray spectral features for X-ray bright single F-type MS stars without significant X-ray outbursts to discuss the coronal properties observationally.
To this end, we conducted spectral analysis on 33 targets with relatively rich X-ray photon statistics 
to extract their spectral characteristics in a steady state.

%
%
%
We successfully extracted physical parameters such as the temperature, the luminosity, and the emission measure, with the absorption column density and the metal abundance fixed at the adopted values.
%
%
%
%
%
%
%
A positive correlation was found in the observed EM--$\bar{kT}$ relationship as seen in late-type stars.
We also found that our single F-type MS stars have a plasma with an EM-weighted temperature of $\lesssim$1 keV and 
an EM of $\lesssim$10$^{53}$ cm$^{-3}$ corresponding to $L_{\rm X}$ of $\lesssim$10$^{30}$~erg~s$^{-1}$.
These results indicate that single F-type and G-type MS stars have similar X-ray coronal properties.
The column EM--$\bar{kT}$ relation, normalized by the surface area of each star, also shows a broadly similar distribution, although some single G-type stars exhibit higher column EM values. This may reflect differences in coronal magnetic activity, potentially associated with different dynamo properties between F- and G-type stars.
%
Our study will strengthen the reliability of extrapolating stellar activity models across these spectral types and provide a consistent basis for studying star--planet interactions.
Note, however, that our results are based on X-ray bright single F-type main-sequence stars selected to ensure sufficiently high photon statistics for reliable spectral analysis and therefore should not necessarily be interpreted as representative of the entire F-type MS star population.

To examine the impact of binarity, 2 single and 9 binary stars were added to our sample by extending the selection conditions. 
As a result, some binary systems broke the upper bounds in EM and $L_{\rm X}$. We also found that the lower bound in the observed EM is likely affected by the selection bias.
%
%
We also studied the relationship between the X-ray activity and the Rossby number, finding that the single stars in our sample are in the X-ray unsaturated regime, while a binary system shows a much higher X-ray luminosity.
%


\begin{acknowledgments}
This study was financially supported by Grants-in-Aid for Scientific Research
(KAKENHI) of the Japanese Society for the Promotion of Science (JSPS, grant Nos. JP18H05438,  JP20K20920, JP21KK0052, and JP22K18274).
S.T. was supported by the JSPS KAKENHI grant Nos. JP18K13579 and JP21H04487.
Based on observations obtained with {\it XMM-Newton}, an ESA science mission with instruments and contributions directly funded by ESA Member States and NASA. 
This research has made use of the SIMBAD database, operated at CDS, Strasbourg, France.
This work has made use of data from the European Space Agency (ESA) mission {\it Gaia} (\url{https://www.cosmos.esa.int/gaia}), processed by the {\it Gaia} Data Processing and Analysis Consortium (DPAC, \url{https://www.cosmos.esa.int/web/gaia/dpac/consortium}). 
Funding for the DPAC has been provided by national institutions, in particular the institutions
participating in the {\it Gaia} Multilateral Agreement.
This paper includes data collected by the {\it TESS} mission, which are publicly available from the Mikulski Archive for Space Telescopes (MAST).
\end{acknowledgments}
\bibliography{references_fstars}{}
\bibliographystyle{aasjournalv7}


\end{document}